\documentclass[prl,aps,showpacs,epsf, twocolumn]{revtex4}

\newcommand{\boldvec}[1]{\mbox{\boldmath$#1$}}

\usepackage{graphicx}
\usepackage{amssymb}

\begin{document}
\title{P-wave Feshbach resonances of ultra-cold $^6$Li}
\author{J.\,Zhang$^{a,b}$, E.\,G.\,M.\,
van Kempen$^c$, T.\,Bourdel$^a$, L.\,Khaykovich$^{a,d}$,
J.\,Cubizolles$^a$, F.\,Chevy$^a$, M.\,Teichmann$^a$,
L.\,Tarruell$^a$, S.\,J.\,J.\,M.\,F.\,Kokkelmans$^{a,c}$,   and
C.\,Salomon$^a$} \affiliation{$^a$ Laboratoire Kastler-Brossel,
ENS, 24 rue Lhomond, 75005 Paris} \affiliation{$^b$ SKLQOQOD,
Institute of Opto-Electronics, Shanxi University, Taiyuan 030006,
P.R. China} \affiliation{$^c$ Eindhoven University of Technology,
P.O.~Box~513, 5600~MB Eindhoven, The Netherlands}
\affiliation{$^d$ Department of Physics, Bar Ilan University,
Ramat Gan 52900, Israel.}

\date{\today}

\begin{abstract}
We report the observation of three p-wave Feshbach resonances of
$^6$Li atoms in the lowest hyperfine state $f=1/2$. The positions
of the resonances are in good agreement with theory. We study the
lifetime of the cloud in the vicinity of the Feshbach resonances
and show that depending on the spin states, 2- or 3-body
mechanisms are at play. In the case of dipolar losses, we observe
a non-trivial temperature dependence that is well explained by a
simple model.
\end{abstract}


\maketitle

In the presence of a  magnetic field, it is possible to obtain a
quasi-degeneracy between the relative energy of two colliding
atoms and that of a weakly bound molecular state. This effect,
known as a Feshbach resonance, is usually associated with the
divergence of the scattering length and is the key ingredient that
led to the recent observation of superfluids from fermion atom
pairs of $^6$Li \cite{Jochim03,Zwirlein03,Bourdel04,Kinast04} and
$^{40}$K \cite{Greiner03}. Up to now these  pairs were formed in
s-wave channels but it is known from condensed matter physics that
fermionic superfluidity can arise through higher angular momentum
pairing: p-wave Cooper pairs have been observed in $^3$He
\cite{RefHe3} and d-wave in high-$T_c$ superconductivity
\cite{RefSupra}. Although  Feshbach resonances involving p or
higher partial waves have been found in cold atom systems
\cite{Vuletic00,Regal03,PWaveBosons}, p-wave atom pairs have never
been directly observed.

In this paper we report  the observation of three narrow p-wave
Feshbach resonances of $^6$Li in the lowest hyperfine state
$f=1/2$. We measure the position of the resonance as well as the
lifetime of the atomic sample for all combinations
$|f=1/2,m_f\rangle+|f=1/2,m'_f\rangle$, henceforth denoted
$(m_f,m'_f)$. We show that the position of the resonances are in
good agreement with theory. In the case of atoms polarized in the
ground state $(1/2,1/2)$, the atom losses are due to 3-body
processes. We show that the temperature dependence of the losses
at resonance cannot be described by the threshold law predicted by
\cite{Esry02} on the basis of the symmetrization principle for
identical particles. In the case of atoms polarized in (-1/2,-1/2)
or that of a mixture (1/2,-1/2), the losses are mainly due to
2-body dipolar losses. These losses show a non trivial temperature
dependence,  that can nevertheless be understood by a  simple
theoretical model with only one adjustable parameter.  In the
(1/2,-1/2) channel, we take advantage of a sharp decrease of the
2-body loss rate below the Feshbach resonance to present a first
evidence for the generation of p-wave molecules.

\begin{figure}
\includegraphics[width=\columnwidth]{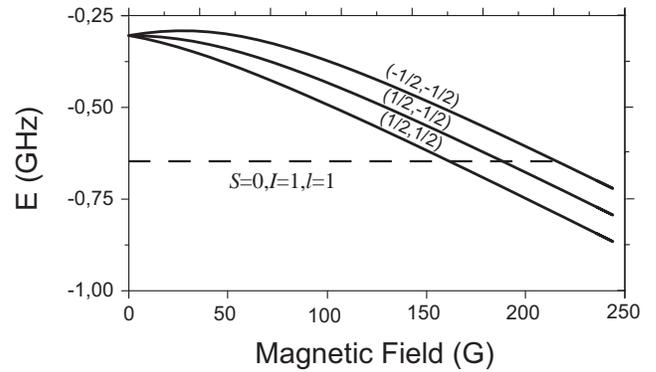}
\caption{Coupled channels calculation of  p-wave binding energies,
which give rise to Feshbach resonances at threshold. The two-atom
states (full line) are indicated by their quantum number
$(m_{f_1},m_{f_2})$, while the bound state (dashed line) is
labelled by the molecular quantum numbers $S,I$, and $l$. }
\label{boundstatediagram}
\end{figure}

The p-wave resonances described in these paper have their origin
in the same singlet ($S=0$) bound state that leads to the s-wave
Feshbach resonances located at $543$~G and $\sim 830$~G.  The
latter has  been used to generate stable molecular Bose-Einstein
condensates \cite{Jochim03, Zwirlein03,Bourdel04,Kinast04}.  In
order to discuss the origin of these resonances, it is useful to
introduce the molecular basis quantum numbers $S,I$, and $l$,
which correspond to the total electron spin $\boldvec S=\boldvec
s_1+\boldvec s_2$, total nuclear spin $\boldvec I=\boldvec
i_1+\boldvec i_2$, and orbital angular momentum $\boldvec l$.
Furthermore, the quantum numbers must fulfill the selection rule

\begin{equation}
S+I+l={\rm even},
\end{equation}
which is a result of the symmetrization requirements of the
two-body wave-function. Since the atomic nuclear spin quantum
numbers are $i_1=i_2=1$, and $S=0$, there are two possibilities
for the total nuclear spin in combination with an s-wave ($l=0$)
collision: $I=0$ and $I=2$. These two states give rise to the two
aforementioned s-wave Feshbach resonances. For p-wave ($l=1$)
collisions only $I=1$ is possible. This bound state may then give
rise to the three p-wave Feshbach resonances of
Fig.~\ref{boundstatediagram}. This threshold state does not suffer
from exchange decay, and is therefore relatively stable. Our
predicted resonance field values $B_{\rm F}$ (Tab. \ref{table1})
result from an analysis which takes into account the most recent
experimental data available for $^6$Li. The calculation has been
performed for all spin channels $(m_f,m'_f)$ and a typical
collision energy of $15~\mu$K. A more detailed analysis will be
published elsewhere~\cite{kempen}.

\begin{table}

\begin{ruledtabular}
\begin{tabular}{ccc}
$(m_{f_1},m_{f_2})$&Theory (G)&Experiment (G)\\
\hline
(1/2,1/2) & 159 & 160.2(6)\\
(1/2,-1/2) &185 & 186.2(6)\\
(-1/2,-1/2) & 215 & 215.2(6)\\
\end{tabular}
\end{ruledtabular}
\caption{\label{table1} Theoretical and experimental values of the
magnetic field $B_{\rm F}$ at the p-wave Feshbach resonance for
$^6$Li atoms in $|f_1=1/2,m_{f_1}\rangle$ and
$|f_1=1/2,m_{f_2}\rangle$. }
\end{table}

Experimentally, we probe these p-wave resonances using the
 setup described in previous
papers~\cite{Bourdel03,Cubizolles03}. After evaporative cooling in
the magnetic trap, we transfer $\sim 5\times 10^5$ atoms of $^6$Li
in $|f=3/2,m_f=3/2\rangle$ in a far-detuned crossed optical trap
at low magnetic field. The maximum power in each arm is $P_{\rm
h}^0=2$~W and $P_{\rm v}^0=3.3$~W in the horizontal and vertical
beam respectively and corresponds to a trap depth of $\sim
80\mu$K. The oscillation frequencies measured by parametric
excitation are respectively $\omega_x=2\pi\times 2.4(2)$~kHz,
$\omega_y=2\pi\times 5.0(3)$~kHz, $\omega_z=2\pi\times
5.5(4)$~kHz, where the $x$ (resp. $y$) direction is chosen along
the horizontal (resp. vertical) beam. A first  radiofrequency (rf)
sweep brings the atoms to $|f=1/2,m_f=1/2\rangle$ and, if
necessary, we perform a second rf transfer to prepare the mixture
$(1/2,-1/2)$ or the pure $(-1/2,-1/2)$. The variable  magnetic
field $B$ is the sum of two independent fields $B_0$ and $B_1$.
$B_0$ offers a wide range of magnetic field while $B_1$ can be
switched off rapidly. After the radio-frequency transfer stage, we
ramp the magnetic field to $B_0\sim 220$~G with $B_1\sim 8$~G in
100~ms. When needed, we reduce in 100~ms the power of the trapping
beams to further cool the atoms. For the coldest samples,
 we obtain at the end of this evaporation sequence $N\sim 10^5$
atoms at a temperature $\sim 5~\mu$K. This corresponds to a ratio
$T/T_F\sim 0.5$, where $k_{\rm B}T_{\rm F}=\hbar
(6N\omega_x\omega_y\omega_z)^{1/3}$ is the Fermi energy of the
system. To reach the Feshbach resonance, we reduce $B_0$ in $4$~ms
to its final value $B_{0,{\rm f}}\sim B_{\rm F}$, near the
Feshbach resonance. At this stage, we abruptly switch off $B_1$ so
that the total magnetic field is now close to resonance. After a
waiting time in the trap $t_{\rm wait}=50$~ms, we switch off the
trapping and the magnetic field and we measure the remaining atom
number after a 0.35~ms time of flight.

\begin{figure}
\includegraphics[width=\columnwidth]{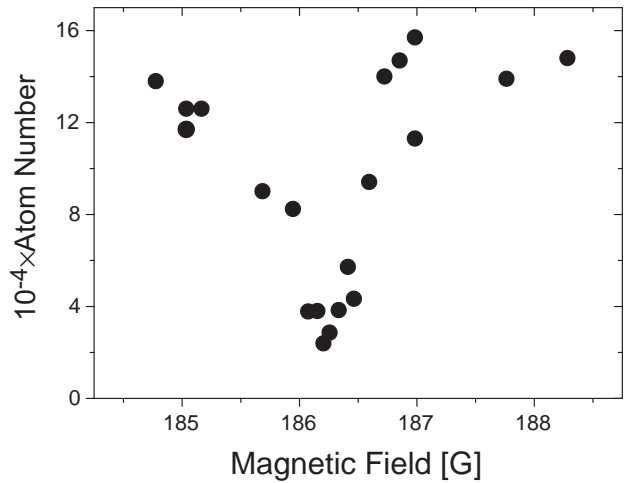}
\caption{Atom number vs. magnetic field $B_{0,{\rm f}}$ after a
50~ms wait for atoms  in the spin mixture $(1/2,-1/2)$ at $T\sim
14\mu$K. The sharp drop close to $B_0\sim 186$~G over a range
$\simeq 0.5$~G is the signature of the p-wave Feshbach resonance
predicted by theory.} \label{Fig1}
\end{figure}

We show in Fig. \ref{Fig1} the dependence of the atom number on
the final value of $B_{0,{\rm f}}$ in the case of the spin mixture
(1/2,-1/2) at a temperature $T\sim 14~\mu$K. As expected from
theory, we observe a sharp drop of the atom number for values of
the magnetic field close to $186$~G. The other two p-wave Feshbach
resonances have a similar loss signature and Tab. \ref{table1}
shows that for all spin channels, the resonance positions are in
good agreement with predictions. Note that in table \ref{table1},
the uncertainty is mainly due to the magnetic field calibration
while the short term stability is $\lesssim 50$~mG.

To evaluate the possibility of keeping p-wave molecules in our
trap, we have studied the lifetime of the gas sample at the three
Feshbach resonances. We have  measured the number $N$ of atoms
remaining in the trap after a variable time $t_{\rm wait}$.
Accounting for 2 and 3-body processes only, $N$ should follow the
rate equation

\begin{equation}
\frac{\dot N}{N}=-G_2\langle n\rangle-L_3\langle n^2\rangle,
\label{RateEqn}
\end{equation}

\noindent where $n$ is the atom density and $\langle
n^a\rangle=\int {\rm d}^3r\, n^{a+1}/N$ ($a=1,2$) is calculated
from the classical Boltzman distribution. In this equation, we can
safely omit one-body losses since the measured decay time is $\sim
100$~ms, much smaller than the one body lifetime $\sim 30$~s.

\begin{figure}
\includegraphics[width=\columnwidth]{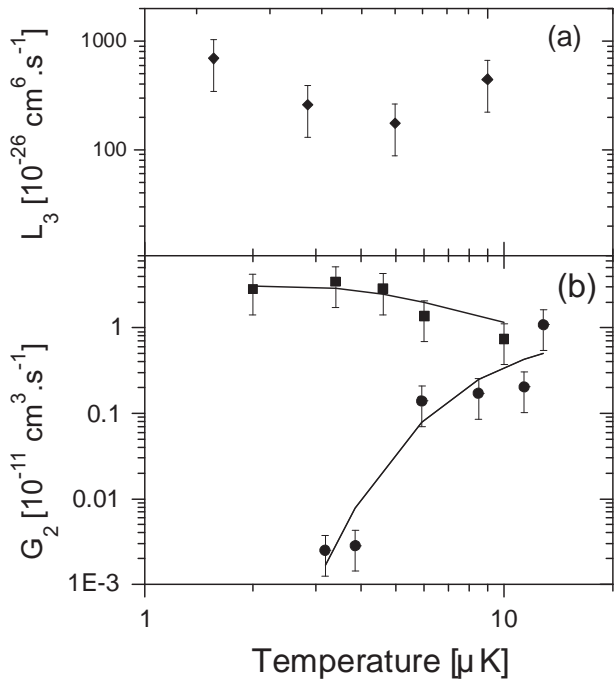}
\caption{Variations of  3-body (a) and 2-body  (b) loss rates vs
temperature at the Feshbach resonance. (a): $\blacklozenge$: atoms
in the Zeeman ground state $|f=1/2,m_f=1/2\rangle$, $B_{0,{\rm
f}}\sim 159$~G. (b):
 $\blacksquare$: atoms polarized in
$|f=1/2,m_f=-1/2\rangle$, $B_{0,{\rm f}}\sim 215$~G. $\bullet$:
mixture $|f=1/2,m_f=1/2\rangle+|f=1/2,m_f=-1/2\rangle$, $B_{0,{\rm
f}}\sim 186$~G. In both cases, the full line is a fit to the data
using prediction of Eq. \ref{Eqn2Body} with the magnetic field as
the only fitting parameter.} \label{Fig3}
\end{figure}

\begin{figure}
\includegraphics[width=\columnwidth]{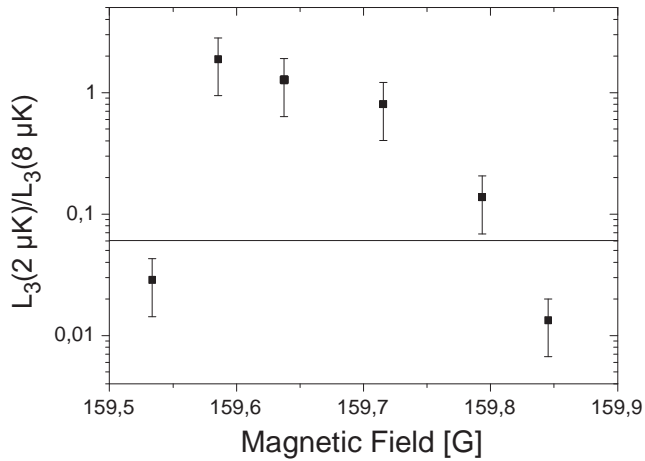} \caption{Ratio $L_3(T=2 \mu{\rm
K})/L_3(T=8~\mu{\rm K})$ of the three body decay rate for two
different temperatures for a gas of atoms polarized in
$|f=1/2,m_f=1/2\rangle$. Full line: threshold law $L_3\sim T^2$.}
\label{Fig4}
\end{figure}

In the (1/2,1/2) channel, we find that 3-body losses are dominant.
 The dependence of $L_3$ with temperature is very weak  (Fig.
\ref{Fig3}.a). A theoretical calculation of the temperature
dependence of 3-body loss rate has been performed in \cite{Esry02}
and it predicts that in the case of indistinguishable fermions
$L_3$ should be proportional to  $T^\lambda$, with $\lambda\ge 2$.
Although this prediction seems in disagreement with our
experimental results,  the analysis of \cite{Esry02} relies on a
Wigner threshold law, {\it i.e.} a perturbative calculation based
on the Fermi golden rule. At the Feshbach resonance where the
scattering cross-section is expected to diverge, this simplified
treatment is not sufficient. This suggests that 3-body processes
must be described by a more refined formalism,
 analogous to the unitary limited treatment of the s-wave
elastic collisions \cite{Esry03}. To confirm this assumption, we
have compared the loss-rates at two given temperatures ($T=2~\mu$K
and $T=8~\mu$K respectively) for various values of the magnetic
field (Fig. \ref{Fig4}). If the threshold law is valid, then the
ratio $L_3(2~\mu {\rm K})/L_3(8~\mu {\rm K})$ should always be
smaller than $(2/8)^2\sim 0.0625$ (full line of Fig. \ref{Fig4}).
As seen before, experimental data show no significant variation of
$L_3$ with temperature near resonance. However, when the magnetic
field is tuned out of resonance we recover a dependence in
agreement with \cite{Esry02}.

In contrast to s-wave Feshbach resonances where dipolar losses are
forbidden in the $f=1/2$ manifold \cite{Dieckmann02},  the losses
at resonance are found to be dominantly 2-body in the (1/2,-1/2)
and (-1/2,-1/2) channels.  The variations of the 2-body loss rate
with temperature are displayed in Fig. \ref{Fig3}.b. The
temperature dependence appears very different in the two cases. We
show now that this is the consequence of a strong sensitivity to
magnetic field detuning from resonance, rather than a specific
property of the states involved. In an extension of the work
presented in \cite{Modele}, we describe inelastic collisions by
two non interacting open channels coupled to a single p-wave
molecular state \cite{FootnoteTicknor}. This model  leads to an
algebra close to the one describing photoassociation phenomena
\cite{Napolitano94} and the 2-body loss rate at energy $E$ is
given by

\begin{equation}
g_2 (E)=\frac{K E}{(E-\delta)^2+\gamma^2/4} \label{Eqn2BodyA}.
\end{equation}

\noindent Here $\delta=\mu (B-B_{\rm F})$ is the detuning to the
Feshbach resonance and $K$, $\mu$ and $\gamma$ are
phenomenological constants depending on the microscopic details of
the potential \cite{FootnoteSign}.
 For each channel, these parameters are estimated from our coupled-channel calculation  (Tab.
\ref{table2}). To compare with experimental data, Eq.
(\ref{Eqn2BodyA}) is averaged over a thermal distribution and for
$\delta>0$ and $\delta\gg\gamma$ we get:

\begin{equation}G_{2}\sim
4\sqrt{\pi}\frac{K}{\gamma}\left(\frac{\delta}{k_{\rm B}
T}\right)^{3/2}{\rm e}^{-\delta/k_{\rm B} T}. \label{Eqn2Body}
\end{equation}

\begin{table}[t]
\begin{ruledtabular}
\begin{tabular}{cccc}
$(m_{f_1},m_{f_2})$&
\begin{tabular}{c}
$K$\\
${\rm cm^3\cdot \mu K\cdot s^{-1}}$
\end{tabular}&
\begin{tabular}{c}
$\gamma$\\
${\rm \mu K}$
\end{tabular}&
\begin{tabular}{c}
$\mu$\\
${\rm \mu K\cdot G^{-1}}$
\end{tabular}
\\
 \hline
(1/2,-1/2)&$1.21\times 10^{-13}$&0.05&117\\
(-1/2,-1/2)&$7.33\times 10^{-13}$&0.08&111\\
\end{tabular}
\end{ruledtabular}
\caption{\label{table2}parameters characterizing the 2-body loss
rates for (1/2,-1/2) and (-1/2,-1/2) spin channels.
 }
\end{table}

 Eqn. \ref{Eqn2Body} is used to fit the data of Fig. \ref{Fig3}.b,
with $B-B_{\rm F}$ as the only fitting parameter. We get a fairly
good agreement if we take $B-B_{\rm F}=0.04$~G (resp. $0.3$~G) for
the (-1/2,-1/2) (resp. (1/2,-1/2)) channel, illustrating the
extreme sensitivity of $G_2$ to detuning and temperature. This
feature was also qualitatively tested by measuring the variations
of $G_2$ with magnetic field at constant temperature. Another
interesting feature of Eqn. \ref{Eqn2Body} is that it predicts
that the width $\delta B$ of the Feshbach resonance, as measured
by atom losses, should scale like $k_{\rm B} T/\mu$. For a typical
temperature $T\sim 15~\mu$K, this yields $\delta B\sim 0.15$~G, in
agreement with the resonance width shown in Fig. \ref{Fig1}.

 From Eq. \ref{Eqn2Body}, we
see that $G_2$ nearly vanishes at $\delta=0$. The thermal average
of (\ref{Eqn2Body}) for $\delta=0$ yields $G_2(\delta=0)\propto K
k_{\rm B} T$. The ratio between the maximum two body loss rate
($\delta=3 k_{\rm B}T/2)$ and that at $\delta=0$ is then $\sim
k_{\rm B} T/\gamma,~\sim 10^2$ for $\sim 10~\mu$K. In the region
$\delta<0$ where we expect to form molecules, we benefit from a
$1/\delta^2$ further reduction of the 2-body losses (see Eqn.
\ref{Eqn2Body}).

\begin{figure}[b]
\includegraphics[width=\columnwidth]{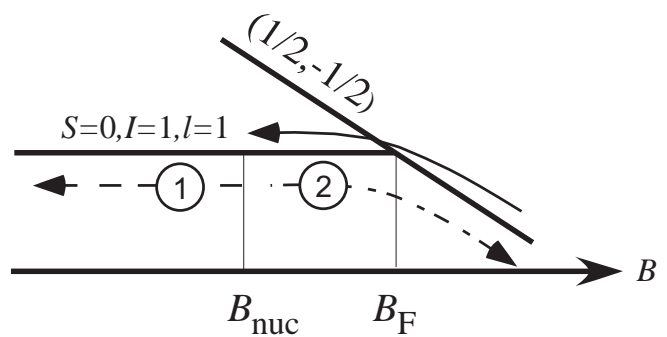}
\caption{Molecules are generated by ramping from a magnetic field
higher than $B_{\rm F}$ to $B_{\rm nuc}<B_{\rm F}$. From there,
two paths are used. In path 1 (dashed line), the magnetic field is
decreased to create tightly bound molecules that will not appear
on absorption images. In  path 2 (dash dotted), the magnetic field
is ramped up across resonance to dissociate the molecules. The
efficiency of the molecule production is simply given by
$(1-N_1/N_2)$ where $N_i$ is the atom number measured after path
$i$.} \label{Fig5}
\end{figure}

 We have
checked the production of molecules in (1/2,-1/2)by using the
scheme presented in \cite{Cubizolles03,Regal03b}. We first
generate molecules  in   $|S=0,I=1,l=1\rangle$ by ramping in 20~ms
the magnetic field from 190~G$>B_{\rm F}$ to $B_{\rm nuc}=185~{\rm
G}<B_{\rm F}$. At this stage, we can follow two paths before
detection (Fig. \ref{Fig5}). Path 1 permits to measure the number
$N_1$ of free atoms: by ramping {\it down} in 2~ms the magnetic
field from 185~G to 176~G, we convert the molecules into deeply
bound molecular states that decay rapidly by 2-body collisions.
Path 2 gives access to the total atom number $N_2$ (free atoms +
atoms bound in p-wave molecules). It consists in ramping {\it up}
the magnetic field in 2~ms from $B_{\rm nuc}$ to 202~G$>B_{\rm F}$
to convert the molecules back into atoms. Since the atoms involved
in molecular states appear only in pictures taken in path 2, the
number of molecules in the trap is $(N_2-N_1)/2$.  In practice,
both sequences are started immediately after reaching $B_{\rm
nuc}$ and we average the data of 25 pictures to compensate for
atom number fluctuations. We then get $N_1=7.1(5)\times 10^4$ and
$N_2=9.1(7)\times 10^4$ which corresponds to a molecule fraction
$1-N_1/N_2=0.2(1)$. Surprisingly, we failed to detect any molecule
signal  when applying the same method to (1/2,1/2) atoms.

Since the dramatic reduction of inelastic losses close to a s-wave
Feshbach resonance  \cite{Petrov03} was a key ingredient to the
recent observation of fermionic superfluids, the formation of
stable atom pairs requires a full understanding of the decay
mechanisms at play close to a p-wave resonance. In this paper we
have shown that in the particular case of 2-body losses, the
maximum losses take place when the detuning is positive. Since
stable dimers are expected to be generated for negative detuning,
dipolar losses should not present a major hindrance to further
studies of p-wave molecules.

We thank Z. Hadzibabic for very helpful discussions.
S.K.~acknowledges supported from the Netherlands Organisation for
Scientific Research (NWO). E.K.~acknowledges support from the
Stichting FOM, which is financially supported by NWO. This work
was supported by CNRS, and Coll\`ege de France. Laboratoire
Kastler Brossel is {\it Unit\'e de recherche de l'\'Ecole Normale
Sup\'erieure et de l'Universit\'e Pierre et Marie Curie,
associ\'ee au CNRS.}

\begin{thebibliography}{99}
\bibitem{Jochim03}S.\,Jochim, {\it et al.}, Science {\bf 302}, 2101 (2003).
\bibitem{Zwirlein03}M.\,W.\,Zwierlein, {\it et al.}, Phys. Rev. Lett.
{\bf 91}, 250401 (2003).
\bibitem{Bourdel04} T. Bourdel {\em et al.}, cond-mat/0403091.
\bibitem{Kinast04}J. Kinast {\em et al.}, Phys. Rev. Lett. {\bf 92}, 150402 (2004).
\bibitem{Greiner03}M.\,Greiner, C.\,A.\,Regal, and D.\,S\,Jin, Nature {\bf 426}, 537 (2003).
\bibitem{RefHe3}D. M. Lee, Rev. Mod. Phys. {\bf 69}, 645 (1997).
\bibitem{RefSupra}C.C Tsuei and J.R. Kirtley, Phys. Rev. Lett. {\bf
85},182 (2000).
\bibitem{Vuletic00} C. Chin
 {\em et al.} Phys. Rev. Lett. {\bf 85}, 2717
(2000).
\bibitem{Regal03}C. A. Regal {\em et al.}
Phys. Rev. Lett. {\bf 90}, 053201 (2003).
\bibitem{PWaveBosons} T. Weber {\em et al.}, Phys. Rev. Lett. {\bf 91}, 123201 (2003).
\bibitem{Esry02} B. D. Esry, C. H. Greene, and H. Suno, Phys. Rev. A {\bf 65}, 010705
(2002).
\bibitem{kempen} E.~G.~M.~van Kempen {\em et al.}, cond-mat/0406722.
\bibitem{Cubizolles03} J.\,Cubizolles, {\it et al}, Phys. Rev. Lett. {\bf 91} 240401 (2003).
\bibitem{Bourdel03}
T.\,Bourdel, {\it et al.}, Phys. Rev. Lett. {\bf 91}, 020402
(2003).
\bibitem{Esry03} H. Suno, B. D. Esry, and C. H. Greene,
Phys. Rev. Lett. 90, 053202 (2003).
\bibitem{Dieckmann02} K. Dieckmann {\em et al.}, Phys. Rev. Lett. {\bf 89}, 203201 (2002).
\bibitem{Modele} M. Holland, {\em et al.}, Phys. Rev. Lett. 87, 120406
(2001).
\bibitem{FootnoteTicknor} We also neglect the splitting between the different
$m_l$ predicted by \cite{ticknor04}.
\bibitem{Napolitano94} R. Napolitano {\em et al.}, Phys. Rev.
Lett. {\bf 73}, 1352 (1994)
\bibitem{ticknor04}C.~Ticknor et al., Phys.~Rev.~A {\bf 69}, 042712 (2004).
\bibitem{FootnoteSign} Note that in our case $\mu$ is positive.
This corresponds to molecular states stable at low field.
\bibitem{Regal03b}C.~A.~Regal {\em et al.}, Nature {\bf 424}, 47 (2003).
\bibitem{Petrov03} D.~S.~Petrov, Phys. rev. A, 67, 010703 (2003).





\end{thebibliography}
\end{document}